\documentclass[12pt]{article}
\usepackage{amsmath, amssymb, geometry}
\usepackage[colorlinks=true, linkcolor=blue, citecolor=teal, urlcolor=magenta, backref=page]{hyperref}
\usepackage{cite}
\usepackage{amsmath, amssymb, geometry}
\usepackage{xcolor}
\usepackage[colorlinks=true, linkcolor=blue, citecolor=teal, urlcolor=magenta, backref=page]{hyperref}
\usepackage{amsmath,amssymb,amsthm,geometry}
\usepackage{cite}

\title{Krein Space Quantization and a Spectral Interpretation of the Riemann $\xi$-Function}
\author{
	M.V. Takook\\
	\small APC, UMR 7164, Universit\'e Paris Cit\'e, 75205 Paris, France\\
	\small Quantum Statetronic, Bagneux, Hauts-de-Seine, France\\
	\small \texttt{takook@apc.in2p3.fr;m.v.takook@gmail.com}
}
\date{\today}

\begin{document}
	\maketitle
	
\begin{abstract}
The invariant two-point function of a scalar field in de Sitter spacetime can be expressed in terms of Legendre functions via Lorentzian harmonic analysis. Using this structure together with the Mehler--Fock transform, we obtain an integral representation of the completed Riemann $\xi$-function in which the Legendre kernel appears naturally. Motivated by this correspondence, we introduce a retarded propagator whose spectral weight is given by the $\xi$-function and analyze it within the framework of Krein space quantization, where sign-indefinite spectral measures are admissible. This construction yields a geometric and spectral interpretation of the $\xi$-function restricted to the critical line and relates the asymptotic spacing of its zeros to a mass--time scaling in de Sitter geometry. The results provide a novel interpretive framework linking de Sitter quantum field theory, harmonic analysis, and analytic number theory.
\end{abstract}

	
\section{Introduction}
	
Quantum Field Theory (QFT) in the de Sitter (dS) ambient-space formalism \cite{ta1} provides a powerful framework for addressing a variety of problems in theoretical physics. Within this setting, one can construct an $N=1$ supersymmetry algebra in de Sitter space \cite{parota} and formulate de Sitter supergravity \cite{taksup}. A finite entropy for quantum fields in the de Sitter universe can also be derived \cite{taen}. When combined with Krein space quantization \cite{gareta00} and quantum metric fluctuations \cite{ta22}, this approach enables the construction of a naturally renormalized QFT. It further allows for the development of a scalar–vector gauge unification scheme \cite{taqg0} and an axiomatic quantum Yang–Mills theory exhibiting both color confinement and a mass gap \cite{taga22,ta231}. Moreover, the ambient-space formalism makes it possible to define de Sitter–invariant $S$-matrix elements \cite{tagahu}. From this observer-independent viewpoint, asymptotic states can be consistently defined in de Sitter space \cite{tagahu}.

Within this framework, and as in a previous work \cite{taqg}, we investigated de Sitter quantum geometry (or, equivalently, a quantum theory of gravity) and reconstructed a complete quantum state of space based on the algebra of elementary field operators. In that analysis, since the divergence of the conformal sector is nonvanishing—implying a breakdown of metric compatibility—we assumed that the quantum conformal sector of gravity could be interpreted as a transition from Riemannian geometry to conformal geometry. Although this assumption has not yet been established experimentally, it does not affect the principal result of that work, namely the construction and renormalization of a complete quantum state of space for quantum gravity \cite{taqg}.
	
In this article, we explore a connection between the completed Riemann $\xi$-function (restricted to the critical line) and the QFT in the dS ambient space formalism. We recall in section \ref{dsqft} the appearance of the Legendre functions of the first and second kind, $P_\lambda^\mu(x)$ and $Q_\lambda^\mu(x)$, as kernel functions in de Sitter QFT \cite{brmo96,brmo2}. We explore the relationship between the Riemann zeta function $\zeta(s)$ and these Legendre functions in section \ref{riem}. In section \ref{zetaret}, using  inverse Fourier–Helgason transforms and Mehler--Fock representation of the Riemann zeta function, we define the retarded propagator in dS space in terms of zeta function, which satisfy all of the nesecery properties. Despite their distinct origins, a nontrivial analytical relationship connects them, primarily through Mellin transforms and the Bros–Fourier–Helgason transform, which appear in integral kernels and retarded propagators. 

In Section~\ref{kallen}, using the Källén–Lehmann representation in de~Sitter space together with Krein space quantization, we present a spectral and geometric interpretation of the nontrivial zeros of the Riemann zeta function in terms of the order of the Legendre function. This order depends continuously on the principal-series parameter~$\nu$, which is interpreted as a mass parameter. The associated mass–time duality is discussed in Section~\ref{ntzamp}, with a brief discussion given in Section~\ref{discussion}. Finally, in Section~\ref{conclusion}, we comment on a possible physical interpretation of the relation between the nontrivial zeros and this mass parameter. In the next section, we briefly recall the notation used throughout the paper.


\section{Notations} \label{notations}

The dS space-time can be identified with the four-dimensional hyperboloid embedded in a five-dimensional Minkowski space-time with the topology $R\times S^3$:
\begin{equation}
	dS_4 = \left\{ x^\alpha \equiv x \in \mathbb{R}^5 \; \big| \;
	x \cdot x = \eta_{\alpha\beta} x^\alpha x^\beta = -H^{-2} \right\},
	\quad \alpha, \beta = 0,1,2,3,4,
\end{equation}
where $\eta_{\alpha\beta} = \mathrm{diag}(1,-1,-1,-1,-1)$, and $H$ denotes a constant parameter analogous to the Hubble constant. The relation between the intrinsic coordinate and the ambient space formalism can be expressed in global coordinates as
\begin{equation} \label{glco}
	 \left\{\begin{array}{clcr} x^0&=H^{-1}\sinh Ht \\
		x^1&=H^{-1}\cosh Ht\sin\chi \cos\theta\\
		x^2&=H^{-1}\cosh Ht\sin\chi \sin\theta\cos\phi \\
		x^3&=H^{-1}\cosh Ht\sin\chi\sin\theta\sin\phi\\
		x^4&=H^{-1}\cosh Ht\cos\chi\;,\\
	\end{array}  \right.\end{equation}
where $(\chi, \theta,\phi)$ are the coordinates of $S^3$. By choosing $\chi\equiv Hr$, the null-curvature limit is obtained:
$$ \lim_{H\longrightarrow 0}x^\alpha=\left( t,r\cos\theta,r\sin\theta\cos\phi,r\sin\theta\sin\phi,\infty\right),$$
which corresponds to the 4-dimensional Minkowski coordinate system.

The analyticity properties of QFT in dS space are described by introducing the complexified de Sitter space-time $dS_4^{(c)}$ \cite{brmo96}:
\begin{align}
	dS_4^{(c)} &= \left\{ z = x + i y \in \mathbb{C}^5 \; \big| \;
	\eta_{\alpha\beta} z^\alpha z^\beta = (z^0)^2 - \vec{z} \cdot \vec{z} - (z^4)^2 = -H^{-2} \right\} \nonumber \\
	&= \left\{ (x, y) \in \mathbb{R}^5 \times \mathbb{R}^5 \; \big| \;
	x^2 - y^2 = -H^{-2}, \; x \cdot y = 0 \right\}.
\end{align}
The forward and backward tubes in $\mathbb{C}^5$ are defined as
\begin{equation}
	T^\pm = \mathbb{R}^5 + i V^\pm,
\end{equation}
where the domains $V^\pm$ are determined by the causal structure of $dS_4$:
\begin{equation}
	V^\pm = \left\{ x \in \mathbb{R}^5 \; \big| \; x^0 \gtrless \sqrt{\| \vec{x} \|^2 + (x^4)^2} \right\}.
\end{equation}
Their respective intersections with $dS_4^{(c)}$ are denoted by
\begin{equation}
	\mathcal{T}^\pm = T^\pm \cap dS_4^{(c)},
\end{equation}
and are called the forward and backward tubes of the complex de Sitter space, respectively. The ``tuboid'' over $dS_4^{(c)} \times dS_4^{(c)}$ is defined as~\cite{brmo96}
\begin{equation}
	\mathcal{T}_{12} = \{ (z, z') \; | \; z \in \mathcal{T}^+, \; z' \in \mathcal{T}^- \}.
\end{equation}
For simplicity, in the next section we set $H = 1$.


\section{de Sitter QFT and Legendre functions} \label{dsqft}

Following Bros and Moschella~\cite{brmo2,brmo96,moschella2024},
the invariant two-point function of a scalar field can be written as
\begin{equation}
	W_\nu (z_1,z_2) = \frac{
		\Gamma\left(\frac{d-1}{2} + i\nu\right)
		\Gamma\left(\frac{d-1}{2} - i\nu\right)
	}
	{2 (2\pi)^{d/2}}
	(\mathcal{Z}^2 - 1)^{-\frac{d-2}{4}}
	P^{-\frac{d-2}{2}}_{-\frac{1}{2}+ i\nu}(\mathcal{Z}),
	\label{wig}
\end{equation}
where $\mathcal{Z}(z_1 , z_2) = z_1 \cdot z_2$, and $d$ is the space–time dimension.
The parameter $\nu$ is the mass parameter corresponding to the principal-series representation of the de Sitter group. The normalization constant is fixed by choosing the Bunch–Davies vacuum state.

For defining the Lorentzian Fourier–Helgason transforms we need the retarded function. To compute the retarded propagator, let us take $x_2$ in the future light cone of the origin $x_o$:
$$
x_o=(0,0,\ldots,0,1), \qquad
x_2(u)=\bigl(\sinh u,\,0,\ldots,0,\,\cosh u\bigr),
\quad
u>0, \quad \mathcal{Z}=-\cosh u.
$$
The retarded propagator for (\(x_2>x_0\)) is therefore given by \cite{moschella2024}:
\begin{eqnarray}
	r_\nu(\mathcal{Z}) &=&
	\frac{i\,\Gamma\!\left(\frac{d-1}{2} + i \nu\right)
		\Gamma\!\left(\frac{d-1}{2} - i \nu\right)}
	{2 (2\pi)^{d/2}}
	(\mathcal{Z}^2-1)^{-\frac{d-2}{4}}
	\left[
	P^{-\frac{d-2}{2}}_{-\frac{1}{2}+ i \nu}(\mathcal{Z}-i\epsilon)
	-
	P^{-\frac{d-2}{2}}_{-\frac{1}{2}+ i \nu}(\mathcal{Z}+i\epsilon)
	\right]
	\nonumber\\[6pt]
	&=&
	\cosh(\pi\nu)
	\frac{
		\Gamma\!\left(\frac{d-1}{2} + i \nu\right)
		\Gamma\!\left(\frac{d-1}{2} - i \nu\right)
	}
	{(2\pi)^{d/2}}
	(\mathcal{Z}^2-1)^{-\frac{d-2}{4}}\,	P^{-\frac{d-2}{2}}_{-\frac{1}{2}+ i \nu}(-\mathcal{Z}).
	\label{retprop}
\end{eqnarray}
The Källén–Lehmann–type representation of retarded functions reads
\begin{equation}\label{kallentr}
	r(\mathcal{Z}) = \int_0^{\infty} d\nu\, \varrho(\nu)\, r_\nu(\mathcal{Z}),
\end{equation}
where $\varrho(\nu)$ is a positive weight function.
This positive measure takes different forms in free and interacting field theories on the dS background and characterizes the type of interaction. Determining its explicit form is one of the key open problems in interacting quantum field theory on dS space.
For further details, see Refs.~\cite{brmo2, brmo96, moschella2024, tagahu, brepmo2}.

For simplicity, let us now restrict to $d = 2$.  
In this case, we have:
\begin{equation}	r_\nu(\mathcal{Z}) =\frac{1}{2}
	\cosh(\pi\nu) P_{-\frac{1}{2}+ i \nu}(-\mathcal{Z}).
	\label{retprop}
\end{equation}
We now introduce the Lorentzian Fourier–Helgason transforms of the retarded function
$r(\cosh u)$, where the support of $r$ is taken into account by setting $x_2(u)=\bigl(\sinh u,\,0,\,\cosh u\bigr)$ with $u \ge 0$. The retarded function is:
\begin{equation} \label{kallntr2}
	r(\mathcal{Z}) = \int_0^{\infty} d\nu\, \varrho(\nu)\, \frac{1}{2}
	\cosh(\pi\nu) P_{-\frac{1}{2}+ i \nu}(\cosh u).
\end{equation} 
The following transforms $G$ and $H$ of $r$ are defined as~\cite{brmo2}:
\begin{equation}
	G(\nu) = \int_{0}^{\infty}
	Q_{-\frac{1}{2} +i\nu}
	(\cosh u)\, r(\cosh u)\, \sinh u\, {\rm d}u,
	\label{qtrans}
\end{equation}
\begin{equation}
	H(\nu) = \int_{0}^{\infty}
	P_{-\frac{1}{2} +i\nu}
	(\cosh u)\, r(\cosh u)\,
	\sinh u\, {\rm d}u,
	\label{ptrans}
\end{equation}
where
$P_{-\frac{1}{2} +i\nu}$
and $Q_{-\frac{1}{2} +i\nu}$
are the Legendre functions of the first and second kind, respectively.  
The classical identity~\cite{bateman1953}
$$
Q_{-\frac{1}{2} +i \nu} -
Q_{-\frac{1}{2} -i \nu} =
-i \pi \tanh(\pi \nu)\, P_{-\frac{1}{2} +i \nu}\;,
$$
implies the relation
\begin{equation}
	H(\nu) = H(-\nu) = \frac{G(\nu)-G(-\nu)}{- i \pi \tanh(\pi \nu)}.
	\label{rel}
\end{equation}

The function $H(\nu)$ is an even function that encodes the spectral content of the invariant perikernel $W(z,z')$. In general, $H(\nu)$ is not automatically positive: its sign depends on the specific form of the underlying the function $W(z,z')$ entering its Legendre transform representation. However, when $W$ is required to be of \emph{positive type}—that is, when the corresponding bilinear form $\int f^*(z)W(z,z')f(z')\,dz\,dz'$ is nonnegative for all test functions $f$—the Fourier–Helgason representation of $W$ implies that the associated spectral weight $\pi H(\nu)/\cosh(\pi\nu)$ must be nonnegative almost everywhere in $\nu$. Thus, the condition of positive definiteness of the kernel $W$ translates directly into the requirement
\begin{equation} \label{positivity}
H(\nu)\ge 0\quad \text{for almost all } \nu\in\mathbb{R},
\end{equation}
which characterizes the admissible class of spectral functions consistent with the positivity (Wightman) condition.


\section{Riemann Zeta Functions} \label{riem}

The Riemann zeta function is defined for $\Re(s) > 1$ by the Dirichlet series
\begin{equation}
	\zeta(s) = \sum_{n=1}^{\infty} \frac{1}{n^{s}}, \qquad \Re(s) > 1.
\end{equation}
It plays a central role in analytic number theory, particularly through its connection with the distribution of prime numbers via the Euler product formula~\cite{edwards2001}.

Riemann introduced a symmetric form of the zeta function by defining
\begin{equation}
	\xi(s)
	= \tfrac{1}{2}\,s(s-1)\,\pi^{-s/2}\,
	\Gamma\!\left(\tfrac{s}{2}\right)\,
	\zeta(s),
\end{equation}
where $\Gamma(s)$ denotes the Gamma function, $\xi(s)$ is the \emph{completed} Riemann zeta function. The completed zeta function satisfies the functional equations
\begin{equation} \label{twoconditions}
	\xi(s) = \xi(1-s), 
	\qquad 
	\xi(s) = [\xi(s^*)]^*,
\end{equation}
where $^*$ denotes complex conjugation. By these relations, if we set $s = \tfrac{1}{2} + i\nu$ with real~$\nu$, the completed zeta function satisfies
\begin{equation}
	\xi\!\left(\tfrac{1}{2}+i\nu\right)
	= \xi\!\left(\tfrac{1}{2}-i\nu\right)
	= \left[\xi\!\left(\tfrac{1}{2}+i\nu\right)\right]^*.
\end{equation}
Therefore, defining 
$$
\Xi(\nu) := \xi\!\left(\tfrac{1}{2}+i\nu\right),
$$
we obtain that $\Xi(\nu)$ is a real and even function of~$\nu$, i.e. $ \Xi(\nu) = \Xi(-\nu)$. The function~$\Xi(\nu)$ is the standard real form of the Riemann $\xi$-function restricted to the critical line. The nontrivial zeros of~$\zeta(s)$ correspond exactly to the real zeros of~$\Xi(\nu)$:
$$
\zeta\!\left(\tfrac{1}{2}+i\nu_n\right)=0
\quad\Longleftrightarrow\quad
\Xi(\nu_n)=0.
$$

It is interesting to note that the conditions~\eqref{twoconditions} are also satisfied by the Legendre function~$P_{-\frac{1}{2}+i\nu}(\cosh u)$, since
$$
P_{-\frac{1}{2}+i\nu}(\cosh u)
= P_{-\frac{1}{2}-i\nu}(\cosh u).
$$
Using the orthogonality relation for Legendre functions~\cite{dlmf,bateman1953},
\begin{equation} \label{ortog1}
	\int_{0}^{\infty}
	P_{-\frac{1}{2}+i\nu}(\cosh u)\,
	P_{-\frac{1}{2}+i\nu'}(\cosh u)\,
	\sinh u\,du
	= \frac{2\pi\,\delta(\nu-\nu')}{\nu\,\tanh(\pi\nu)},
\end{equation}
together with its inversion formula,
\begin{equation} \label{ortog2}
	\frac{1}{\pi}\int_{0}^{\infty}
	P_{-\frac{1}{2}+i\nu}(\cosh u)\,
	P_{-\frac{1}{2}+i\nu}(\cosh v)\,
	\nu\tanh(\pi\nu)\,d\nu
	=\frac{\delta(u-v)}{\sinh u},
\end{equation}
one obtains a Mehler--Fock representation of the Riemann $\xi$-function in terms of Legendre functions:
\begin{equation}\label{eq:xi_P}
	\Xi(\nu)
	=\int_{0}^{\infty}
	P_{-\frac{1}{2}+i\nu}(\cosh u)\,\Phi(u)\,
	\sinh u\,du,
\end{equation}
where $\Phi(u)$ denotes the corresponding spectral amplitude. The corresponding inverse transform gives an explicit integral representation for $\Phi(u)$:
\begin{equation}\label{phiu}
	\Phi(u)=\frac{2}{\pi}\int_{0}^{\infty}
	\Xi(\nu)\,
	P_{-\frac{1}{2}+i\nu}(\cosh u)\,
	\nu\tanh(\pi\nu)\,d\nu.
\end{equation}
Hence, $\Xi(\nu)$ is precisely the \emph{Mehler--Fock transform} of the function $\Phi(u)$, which can be recovered from $\Xi(\nu)$ via the standard inversion weight $\nu\tanh(\pi\nu)$.

\section{Zeta function and Retarded propagator}\label{zetaret}

Drawing inspiration from Eqs.~\eqref{eq:xi_P} and~\eqref{ptrans}, we introduce the identification of the Mehler–Fock spectral amplitude with a de Sitter retarded response as a \emph{structural ansatz}. Specifically, we set $\Phi(u)\equiv \mathcal{R}(\cosh u)$ and interpret $\mathcal{R}$ as a candidate invariant retarded propagator whose Fourier–Helgason data are given by $\Xi(\nu)$. This identification is not derived from a microscopic interacting model; rather, it is motivated by the common harmonic-analysis structure of the two representations: both Eq.~\eqref{ptrans}, which encodes retarded Fourier–Helgason data, and Eq.~\eqref{eq:xi_P}, which provides the Mehler–Fock representation of $\Xi$, involve the same Legendre kernel. Promoting this ansatz to a theorem would require the construction of an explicit de Sitter-invariant dynamical model whose retarded two-point function reproduces $\Phi(u)$, which is beyond the scope of the present paper.
Then Eq.~\eqref{phiu} can be rewritten as:
\begin{equation}
	\mathcal{R}(\cosh u)
	\equiv \frac{2}{\pi} \int_{0}^{\infty}
	\Xi(\nu)\, P_{-\frac{1}{2}+i\nu}(\cosh u)\, \nu\, \tanh(\pi\nu)\, {\rm d}\nu.
	\label{rpwz}
\end{equation}
We now show that the integral representation $ \mathcal{R}(\cosh u)$ in~\eqref{rpwz}, with Riemann zeta function $\Xi(\nu)$, satisfies all the properties required for a retarded propagator on dS one–sheeted hyperboloid.

\paragraph{1. Causality.}
In the Lorentzian de~Sitter setting, ``retarded'' is a support property in the variable $u$, corresponding to the future light-cone condition (see Eqs.~\eqref{retprop} and \eqref{kallntr2}). Equation~\eqref{rpwz} defines $\mathcal{R}(\cosh u)$ using the same Lorentzian Fourier--Helgason/Mehler--Fock kernel in $u$ as in the de~Sitter retarded representation. The superposition over $\nu$ does not enlarge the support in $u$: replacing a positive spectral weight by the sign-indefinite function $\Xi(\nu)$ modifies the spectral sign structure but does not generate support outside the future domain. Hence $\mathcal{R}(\cosh u)$ retains retarded (future-supported) behavior in the same sense as in Ref.~\cite{brmo2}.

\paragraph{2. Correct Fourier--Helgason Data.}
Using the ortogonality relation \eqref{ortog1} in Eq.~\eqref{rpwz}, we obtain
\begin{equation} \label{zetalog}
 \Xi(\nu) = \int_{0}^{\infty} P_{-\frac{1}{2}+i\nu}(\cosh u)\,\mathcal{R}(\cosh u)\,\sinh u\,{\rm d}u,
\end{equation}
which is the standard retarded/advanced Fourier--Helgason structure (see Ref.~\cite{brmo2} for details).

\paragraph{3. Regularity / Integrability.}
Proposition~15 of \cite{brmo2} requires
$
e^{u/2}\,r(\cosh u) \in L^{1}(\mathbb{R}_{+}, {\rm d}u),
$
which is valid in this case since the Legendre function in two cases is the same. On the critical line, $\Xi(\nu)$ decays asymptotically as $e^{-\pi|\nu|/4}$
(according to Stirling’s expansion of the Gamma function in the completed $\xi(s)$-function), whereas the Mehler–Fock kernel contributes only polynomial factors in $|\nu|$. Thus the inversion integral defines a real function $\mathcal{R}$ decaying faster than $e^{-u/2}$,
which satisfies the required condition.

\paragraph{4. Reality.}
Since $\Xi(\nu)$ is real and even for real $\nu$,
and $P_{-\frac{1}{2}+i\nu}(\cosh u)$ is real for $u \ge 0$,
it follows that $\mathcal{R}(\cosh u)$ is also real.

In summary, the function $ \mathcal{R}(\cosh u)$ defined by equation \eqref{rpwz} yields a genuine retarded propagator in the sense of \cite{brmo2}:
it possesses the causal property, produces $\Xi(\nu)$ under the Fourier–Helgason transform, and satisfies the required decay and reality conditions. However, it differs from the equation \eqref{ptrans}. Unlike $H(\nu)$, which is positive definite, the function $\Xi(\nu)$ is not positive, and its sign change corresponds to the nontrivial zero of the Riemann zeta function. We will address this point in the next section.


\section{Krein Space quantization and indefinite spectral measure} \label{kallen}

The Källén–Lehmann spectral representation provides a powerful framework for expressing the two-point correlation functions of an interacting field in terms of a spectral measure $\varrho(\nu)$, which encodes the dynamical content of the field and its interactions (see Eq.~\eqref{kallentr}), as well as in terms of the two-point correlation functions of the corresponding free field. The spectral density $\varrho(\nu)$ is usually nonnegative, reflecting the positivity of the Hilbert-space inner product. In gauge-invariant theories \cite{stroch}, as well as in Krein-space quantization \cite{gareta00}, however, maintaining full covariance requires quantization in an indefinite inner-product space that contains both positive- and negative-norm modes. In such cases, the positivity condition on $\varrho(\nu)$ is no longer required.

A Krein space is a complex linear space with an indefinite inner product. It can be decomposed into an orthogonal direct sum of two subspaces: one with a positive-definite inner product and the other with a negative-definite inner product \cite{gareta00}:
\begin{equation} \label{onepks} \mathcal{K} \equiv \mathcal{H} \oplus \mathcal{H}^*\,,
\end{equation}
where $\mathcal{H}$ is a Hilbert space. In this setting, the resulting two-point function implies that the corresponding spectral measure~$\varrho(\nu)$ becomes sign-indefinite. This loss of spectral positivity is not pathological but rather a necessary feature ensuring de~Sitter invariance as well as infrared and ultraviolet regularity, particularly for massless minimally coupled scalar fields and in quantum gravity \cite{taqg}.

Comparing Eqs.~\eqref{kallntr2} and~\eqref{rpwz} yields the spectral weight in the form
\begin{equation}
	\label{kreinspectral}
	\varrho(\nu)
	= \frac{4}{\pi}\,
	\frac{\nu\,\tanh(\pi\nu)}{\cosh(\pi\nu)}\,
	\Xi(\nu).
\end{equation}
This expression is analogous to Eq.~(106) of Ref.~\cite{brmo2}, with the function $H(\nu)$ replaced by $\Xi(\nu)$. The sign variations of the factor~$\Xi(\nu)$ in the spectral density can naturally be associated with the Krein-space structure, linking the analytic behavior of~$\Xi(\nu)$ to the alternating contributions from positive- and negative-norm sectors. In this context, a spectral function of the form~\eqref{kreinspectral} can be consistently interpreted as a \emph{Krein-type spectral density}. The possible sign changes of~$\Xi(\nu)$ reflect the indefinite structure of the underlying representation space.

The sign changes of~$\Xi(\nu)$ determine its zeros, which, from Eq.~\eqref{zetalog}, can be related to the zeros of the Legendre function~$P_{-\frac12 + i\nu}(x)$ in terms of parameter $\nu$, since the integral representation~\eqref{zetalog} is regular. The order of this Legendre function depends continuously on the principal-series parameter~$\nu$. Its zeros for the fixed variable $x=\cosh u$ vary smoothly with~$\nu$, reflecting the continuous family of modes associated with the principal-series representations. This motivates a spectral representation in which two-point functions are expressed as superpositions over the mass parameter~$\nu$, yielding in the de~Sitter setting, which results from the Källén--Lehmann decomposition. In the spherical transform~\eqref{ptrans} or~\eqref{zetalog}, the kernel $P_{-\frac12+i\nu}(\cosh u)$ acts as the spherical Fourier kernel on the rank-one symmetric space
\(\mathrm{dS}_2 \simeq {\rm SO}_0(1,2)/{\rm SO}(2)\),
so that $H(\nu)$ or $\Xi(\nu)$ represents the principal-series projection of the retarded functions.

The appearance of the completed Riemann zeta function~$\Xi(\nu)$ in the spectral weight~$\varrho(\nu)$ suggests a deep interplay between the analytic structure of number theory and the spectral geometry of quantum fields in dS spacetime. In the present framework, the zeros and oscillatory behavior of~$\Xi(\nu)$ reflect the alternating contributions of positive- and negative-norm sectors inherent to Krein-space quantization. These sign variations may thus be viewed as encoding a nontrivial spectral pattern, analogous to the resonance structure of normal modes in~$P_{-\frac12 + i\nu}(x)$. From this perspective, the Källén--Lehmann measure~$\varrho(\nu)$ not only determines the causal response of the field but also suggests an underlying arithmetic or spectral structure, thereby motivating a possible connection between analytic number theory and quantum-field dynamics in a de~Sitter background. In particular, the zeros of integral representation of~$P_{-\frac12 + i\nu}(x)$ or of the completed Riemann zeta function~$\Xi(\nu)$ may correspond to special resonant configurations or effective mass parameters, thereby providing a possible physical insight into the spectral organization of the theory.

Within the Krein-space spectral formulation, values of the parameter $\nu$ for which $\varrho(\nu)=0$ correspond to null contributions in the spectral decomposition: the associated mode carries no propagating weight in the retarded response. This vanishing reflects cancellations inherent to the indefinite inner-product structure of Krein space, rather than the presence of singularities, instabilities, or physical excitations.

From this geometric perspective, the interpretation of the zeros is structurally distinct from operator-based approaches to the Riemann zeros, such as the Hilbert--Pólya conjecture \cite{Connes99} and the Berry--Keating proposal \cite{BerryKeating99}. In the present formulation, this distinction is conceptual rather than predictive: the framework does not introduce a quantum Hamiltonian or eigenvalue problem, but instead relates the zeros to null spectral contributions in a dS causal and harmonic-analysis setting, whose potential observable consequences would require the construction of an explicit dynamical model.


\section{Nontrivial zeros and mass parameter} \label{ntzamp}

It is important to note that taking the limits $H \to 0$ (the null-curvature or Minkowskian limit) and $\nu \to \infty$ such that $H\nu$ remains finite yields the concept of mass in Minkowski space \cite{brmo96}:
\begin{equation} \label{maspa}
\lim_{\substack{H \to 0 \\ \nu \to \infty}} H\nu  = m.
\end{equation}
We then consider the zeros of the Legendre function in the limit $\nu \to \infty$. For real $\nu$, the conical–Legendre function $P_{-\tfrac12+i\nu}(x)$ is real-valued, even in $\nu$, and has an infinite sequence of simple real zeros $\{\pm \nu_n\}_{n\ge 0}$ in $\nu$. Let $x>1$ be fixed and write $ x=\cosh u$. Their behavior is governed by the standard large-\(\nu\) asymptotic (uniform for fixed \(u >0\)) \cite{dlmf}:
\begin{equation} \label{nulimit}
\lim_{\nu\to\infty } P_{-\frac12+i\nu}(\cosh u)\sim \sqrt{\frac{2}{\pi \nu\sinh u}}\,
\cos\,\Bigl(\nu u-\frac{\pi}{4}\Bigr).
\end{equation}
Consequently, the zeros in \(\nu\) for fixed \(x>1\) satisfy (to leading order)
$$
\nu_n\,u-\frac{\pi}{4}\approx \Bigl(n+\tfrac12\Bigr)\pi
\quad\Longrightarrow\quad
\displaystyle
	\nu_n \approx \frac{\bigl(n+\tfrac34\bigr)\pi}{u}
	= \frac{\bigl(n+\tfrac34\bigr)\pi}{\operatorname{arcosh}x},
\qquad n=0,1,2,\dots
$$
with nearly constant spacing
$$
\nu_{n+1}-\nu_n \approx \frac{\pi}{u} = \frac{\pi}{\operatorname{arcosh}x}.
$$

Now we consider the zero of the integral representation \eqref{zetalog}, with $\mathcal{R}$ regular so that the integral converges. Using the conical–Legendre large–$\nu$ asymptotic (uniform for fixed $u>0$) \eqref{nulimit}, we obtain:
$$
\lim_{\nu\to\infty }\Xi(\nu)=\sqrt{\frac{2}{\pi\nu}}\int_{0}^{\infty}
\underbrace{\sqrt{\sinh u}\,\mathcal{R}(\cosh u)}_{=:g(u)}
\cos\,\Bigl(\nu u-\frac{\pi}{4}\Bigr)\,\mathrm{d}u
+O\!\left(\nu^{-3/2}\right)\equiv I(\nu).
$$
The asymptotic zeros depend on the right–hand function of $g$. We consider two simple cases.
\paragraph{1) Compact support.}  
If $\mathcal{R}(\cosh u)=0$ for $u>u_c$ with some $u_c>0$ and $g(u_c)\neq 0$, an integration–by–parts (endpoint) analysis gives
$$
I(\nu)=\sqrt{\frac{2}{\pi}}\,
\frac{g(u_c)}{\nu^{3/2}}\,
\sin\,\Bigl(\nu u_c-\frac{\pi}{4}\Bigr)
+O\!\left(\nu^{-5/2}\right),
$$
hence the zeros $\{\nu_n\; ; n=0,1,2,\dots\}$ satisfy
$$
	\nu_n \;\approx\; \frac{\bigl(n+\tfrac14\bigr)\pi}{u_c},
	\qquad \;\nu_{n+1}-\nu_n\approx \frac{\pi}{u_c}.
$$
\paragraph{2) Fast decay.}  
If $\mathcal{R}$ decays smoothly as $u\to\infty$ (no finite $u_c$), there is no universal closed form for the zeros: $I(\nu)$ is essentially a cosine transform of $g$, so its large–$\nu$ behavior and zeros are governed by the far-right tail of $g$. A practical rule is:
\begin{equation} \label{spavt}
\nu_n \;\approx\; \frac{(n+\tfrac14)\pi}{u_{\mathrm{eff}}},
\qquad 
\nu_{n+1}-\nu_n \approx \frac{\pi}{u_{\mathrm{eff}}},
\end{equation}
where $u_{\mathrm{eff}}$ is an effective decay scale determined by the region where $g(u)$ becomes negligible. The endpoint $u=0$ does not contribute at leading order because $g(0)=0$. If $g(u)=0$ but $g^{(k)}(u)\neq0$ for some $k\ge 1$, the zero locations remain the same while the amplitude gains an additional factor $\nu^{-k}$.

Using the Riemann-von~Mangoldt zero counting formula for non-trivial zeros of the Riemann zeta function,
$$
N(\nu)
= \frac{\nu}{2\pi}\log\,\Bigl(\frac{\nu}{2\pi}\Bigr)
- \frac{\nu}{2\pi}
+ \frac{7}{8}
+ O(\log \nu),
$$
the average spacing between zeros in the limit $\nu \to \infty $ is obtained as \cite{edwards2001}:
\begin{equation} \label{spavr}
		\nu_{n+1}-\nu_n  \sim \frac{2\pi}{\log(\nu_n/2\pi)}.
\end{equation}
By comparing equation \eqref{spavr} and \eqref{spavt}, we obtain:
\begin{equation} \label{mati}
	u_{\mathrm{eff}} \approx \frac{1}{2}\,\log\!\left(\frac{\nu_n}{2\pi}\right).
\end{equation}

In the intrinsic global coordinate system \eqref{glco}, the parameter $u$ can be regarded as the physical time variable for an observer at rest at the spatial origin. With this interpretation, the behavior of the nontrivial zeros of the Riemann zeta function at large $\nu$ is governed by the late-time decay of the underlying function $g(u)$. Physically $\nu$ play the role of the mass and equation 
\eqref{mati} can be interpreted as the mass-time relation in dS geometry, indicating that each mode with mass parameter $\nu$ is naturally associated with the effective time $u$ determined by this relation. This can be viewed as a form of mass–time duality: heavier modes correspond to longer characteristic time scales. It provides a direct link between the mass scale of a mode and the effective time at which it contributes to the dynamics. The logarithmic dependence reflects the exponential structure of dS expansion, whereby mass thresholds induce shifts in the physical time coordinate rather than linear scalings.


\section{Discussion}\label{discussion}

The results presented in this work are primarily interpretive and structural in nature. In particular, the identification $\Phi(u)\equiv \mathcal{R}(\cosh u)$ is introduced as a motivated ansatz based on the common Mehler--Fock and Legendre-kernel structure, rather than derived from a microscopic interacting quantum field theory on dS space. Establishing such a derivation would require the construction of an explicit dS-invariant dynamical model whose causal response reproduces the spectral data of the completed Riemann $\xi$-function.

Furthermore, while the large-$\nu$ asymptotic analysis of the Legendre kernel provides insight into the spacing and scaling behavior of the nontrivial zeros, the present framework does not constrain their exact locations. The analysis applies to the $\xi$-function restricted to the critical line and assumes its standard analytic properties.

From the perspective of analytic number theory, the present framework does not provide a mechanism to fix the exact location of the nontrivial zeros or to establish their confinement to the critical line. Achieving such control would require identifying a dynamical principle or symmetry capable of determining the spectral support of $\Xi(\nu)$ beyond its asymptotic behavior. Whether a dS--invariant quantum field theoretic model can supply such a principle remains an open question and a subject for future investigation.

The Krein-space formulation adopted here permits sign-indefinite spectral measures and ensures covariance and regularity, but it does not imply a one-to-one correspondence between individual zeros of $\Xi(\nu)$ and physical states. Despite these limitations, the framework suggests a novel geometric perspective in which spectral parameters acquire a spacetime interpretation. With these limitations in mind, we summarize the main results and outline possible directions for future work in the following Conclusion section.

\section{Conclusion and Outlook} \label{conclusion}

We have shown that the completed Riemann $\xi$-function $\Xi(\nu)$ admits a representation in terms of Legendre functions via the Mehler--Fock transformation. Within the framework of dS quantum field theory and Krein space quantization, this representation identifies $\Xi(\nu)$ with the spectral component of a retarded response, thereby clarifying its relation to the causal and harmonic-analysis structure of de~Sitter spacetime. From this perspective, the spectral parameter $\nu$ acquires a natural geometric interpretation associated with principal-series representations of the dS group.

This correspondence suggests a geometric viewpoint in which aspects of analytic number theory are reflected in the spectral organization of quantum fields on curved spacetime. In particular, the complex parameter $s$, appearing in discussions of the Riemann Hypothesis as $s=\tfrac{1}{2}\pm i\nu$, may be viewed as encoding a mass-like parameter in dS harmonic analysis. Future work may explore whether this framework can be embedded into explicit dynamical models, whether analogous constructions arise in other symmetric spacetimes, and whether such settings provide further insight into the interplay between spectral geometry, causal structure, and number-theoretic functions.

\section*{Acknowledgements}
The author dedicates this work to the memory of Jacques Bros, whose pioneering contributions to harmonic analysis in the de~Sitter ambient space formalism continue to illuminate this rich field of research. 



\end{document}